\documentclass[prl,aps,twocolumn,showpacs,superscriptaddress]{revtex4-1}
\usepackage{amssymb}
\usepackage{amsmath}
\usepackage[version=4]{mhchem}
\usepackage{hyperref}
\hypersetup{colorlinks=true, citecolor=blue, filecolor= black, linkcolor= blue, urlcolor= blue}
\usepackage{graphicx}
\usepackage{dcolumn}
\usepackage{bm}
\usepackage{float}
\usepackage{braket}
\begin{document}

\title{Unusual phase transition of layer-stacked borophene under pressure}

\author{Xiao-Ji Weng}
\affiliation{Key Laboratory of Weak-Light Nonlinear Photonics and School of Physics, Nankai University, Tianjin 300071, China}
\affiliation{State Key Laboratory of Metastable Materials Science and Technology, School of Science, Center for High Pressure Science, Yanshan University, Qinhuangdao 066004, China}

\author{QuanSheng Wu}
\affiliation{Institute of Physics, $\acute{E}$cole Polytechnique F$\acute{e}$d$\acute{e}$rale de Lausanne (EPFL), CH-1015 Lausanne, Switzerland}
\affiliation{National Centre for Computational Design and Discovery of Novel Materials MARVEL, Ecole Polytechnique F$\acute{e}$d$\acute{e}$rale de Lausanne (EPFL), CH-1015 Lausanne, Switzerland}

\author{Xi Shao}
\affiliation{State Key Laboratory of Metastable Materials Science and Technology, School of Science, Center for High Pressure Science, Yanshan University, Qinhuangdao 066004, China}

\author{Oleg V. Yazyev}
\affiliation{Institute of Physics, $\acute{E}$cole Polytechnique F$\acute{e}$d$\acute{e}$rale de Lausanne (EPFL), CH-1015 Lausanne, Switzerland}
\affiliation{National Centre for Computational Design and Discovery of Novel Materials MARVEL, Ecole Polytechnique F$\acute{e}$d$\acute{e}$rale de Lausanne (EPFL), CH-1015 Lausanne, Switzerland}

\author{Xin-Ling He}
\affiliation{Key Laboratory of Weak-Light Nonlinear Photonics and School of Physics, Nankai University, Tianjin 300071, China}

\author{Xiao Dong}
\affiliation{Key Laboratory of Weak-Light Nonlinear Photonics and School of Physics, Nankai University, Tianjin 300071, China}

\author{Hui-Tian Wang}
\affiliation{National Laboratory of Solid State Microstructures and Collaborative Innovation Center of Advanced Microstructures, Nanjing University, Nanjing 210093, China}

\author{Xiang-Feng Zhou}
\email{xfzhou@nankai.edu.cn}
\email{zxf888@163.com}
\affiliation{Key Laboratory of Weak-Light Nonlinear Photonics and School of Physics, Nankai University, Tianjin 300071, China}
\affiliation{State Key Laboratory of Metastable Materials Science and Technology, School of Science, Center for High Pressure Science, Yanshan University, Qinhuangdao 066004, China}

\author{Yongjun Tian}
\affiliation{State Key Laboratory of Metastable Materials Science and Technology, School of Science, Center for High Pressure Science, Yanshan University, Qinhuangdao 066004, China}


\begin{abstract}
 The 8-$Pmmn$ borophene, a boron analogue of graphene, hosts tilted and anisotropic massless Dirac fermion quasiparticles owing to the presence of the distorted graphene-like sublattice. First-principles calculations show that the stacked 8-$Pmmn$ borophene is transformed into the fused three-dimensional borophene under pressure, being accompanied by the partially bond-breaking and bond-reforming. Strikingly, the fused 8-$Pmmn$ borophene inherits the Dirac band dispersion resulting in an unusual semimetal-semimetal transition. A simple tight-binding model derived from graphene qualitatively reveals the underlying physics due to the maximum preservation of graphene-like substructure after the phase transition, which contrasts greatly to the transformation of graphite into diamond associated with the semimetal-insulator transition.
\end{abstract}



\maketitle

Graphene, a monolayer material exfoliated from graphite, has attracted immense interest partially owing to its remarkable thermal stability, superior mechanical strength and unique Dirac-semimetallic band structure \cite{R01,R02,R03}. Graphene shows a significant potential for applications in the fields of electronics, batteries, sensors, structural materials and more. Boron, the element that neighbors carbon in the second row of the periodic table and forms a nonmetallic solid, has also been proved to own its two-dimension (2D) form, termed borophene \cite{R04,R05,R06}. Borophene shares a lot in common with graphene, but has its own uniqueness due to the complex multicenter bonds. Both theoretical simulations and experiments extensively explored the diverse geometrical configurations (i.e., planar, quasi-planar, or multilayer structures) and versatile properties (i.e., superconductivity, magnetism, or negative Poisson$^\prime$s ratio) in various borophenes \cite{R04,R05,R06,R07,R08}. For instance, the 8-$Pmmn$ borophene, featured by an 8-atom orthorhombic unit cell with the $Pmmn$ symmetry \cite{R09}, is composed of distorted graphene-like boron sheet and attached boron chains. More specifically, the boron chains attach above and below the distorted graphene-like sheet. Each B-B pair along the chain direction donates two electrons to the distorted hexagonal lattice, satisfying the isoelectronic state of graphene and resulting in the tilted and anisotropic Dirac cone at the Fermi level ($E$$_{\rm F}$), which is quite different from that of graphene (isotropic Dirac cone) \cite{R10,R11}. The maximum Fermi velocity of 8-$Pmmn$ borophene along the chain direction is nearly 1.5 times that of graphene \cite{R09}. Such unique band structure stimulated further study of the electronic properties of 8-$Pmmn$ borophene, such as anisotropic plasmons, magnetotransport properties, anomalous Klein tunneling, valley-dependent electron retroreflection, Veselago focusing, Ruderman-Kittel-Kasuya-Yosida exchange interaction, metal-insulator transition, and the photoinduced Hall effect \cite{R12,R13,R14,R15,R16,R17,R18,R19,R20,R21}. Most recently, a variety of exotic electronic states, including flat-band phases, superconductivity, magnetism, and the Chern insulators phases, were discovered moir$\acute{\rm e}$ superlattice systems based on multilayer graphene or other nodal-line semimetals \cite{R22,R23,R24,R25,R26,R27,R28,R29,R30,R31}. These new discoveries made a very strong impact in condensed matter physics and beyond. Obviously, it is interesting to investigate the ambient pressure bulk stacked borophenes as well as their high-pressure phases.

In this work, boron polymorphs derived from 8-$Pmmn$ borophene have been studied using the projector augmented wave (PAW) method \cite{R32} as implemented in the VASP code \cite{R33}. The exchange correlation energy was treated within the generalized gradient approximation (GGA) using the functional of Perdew, Burke, and Ernzerhof (PBE) \cite{R34}. The plane-wave energy cutoff of 500 eV, energy convergence criterion of 10$^{-7}$ eV, force criterion of 0.003 eV/{\AA}, and $k$-point resolution of 2$\pi$ $\times$ 0.04~\AA$^{-1}$ were employed for the density functional theory (DFT) calculation, which showed excellent convergence of the total energy, stress tensors and lattice parameters. To examine the energetic and dynamical stability, the semiempirical dispersion-correction method (DFT-D3) was applied to the various boron allotropes \cite{R35}. Phonon spectrum was calculated by the \textsc{phonopy} code \cite{R36} within the density functional perturbation theory (DFPT) framework. Moreover, \textit{ab initio} molecular dynamics (AIMD) simulations within the canonical ensemble (NVT) were preformed using a $2 \times 2 \times 2$ supercell and Nos\'e-Hoover thermostat \cite{R37}.


\begin{figure}[t]\label{f1}
\begin{center}
\includegraphics[width=8.0cm]{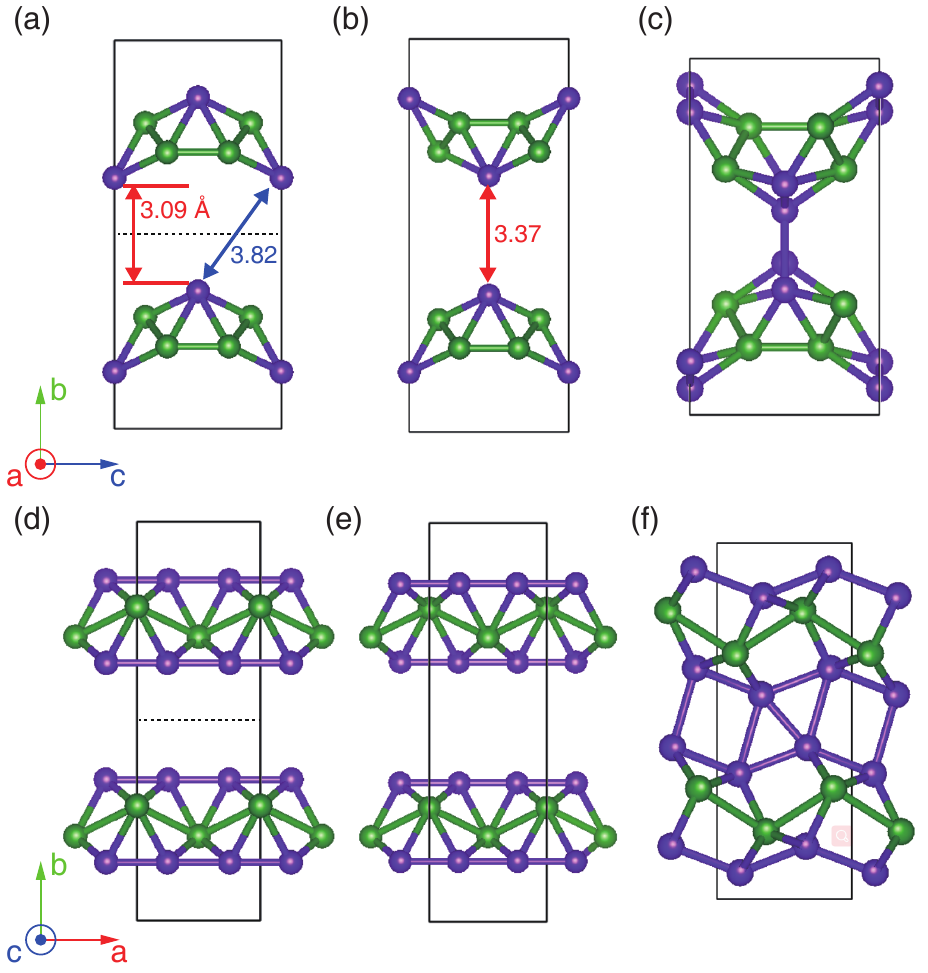}
\caption{%
(a)-(c) Projection along the [100] direction of the AA-8-$Pmnm$ (1$\times$2$\times$1 supercell), AB-16-$Amam$ (conventional cell), and the fused AB-16-$Pnnm$ phase. (d)-(f) Projection along the [001] direction of the AA-8-$Pmnm$, AB-16-$Amam$, and the fused AB-16-$Pnnm$ phases. The distorted graphene-like sublattice is colored in green while the attached boron chains are colored in purple.}
\end{center}
\end{figure}

Two layer-stacked borophene configurations are constructed from the parent 8-$Pmnm$ structure, and designated as the AA-8-$Pmnm$ and AB-16-$Amam$ borophenes. Two sublattices of the 8-$Pmnm$ borophene are illustrated in Fig. \hyperref[f1]{1} using different colors. The AA-8-$Pmnm$ configuration was constructed from the 8-$Pmnm$ borophene without introducing any lateral shift in the (010) plane, while there is a shift of half of lattice parameter \textbf{\emph{c}} in the AB-16-$Amam$ phase. The lattice parameters and atomic positions of the stacked borophenes compared with other related boron allotropes are listed in Table \hyperref[t1]{I}. The lattice parameters $a$ and $c$ of AA-8-$Pmnm$ and AB-16-$Amam$ borophenes are almost the same with those of the 8-$Pmmn$ structure. The B-B bond lengths of these two stacked borophenes are ranging from 1.61 to 1.91 {\AA}. To investigate the interlayer distance of the stacked borophenes, the vdW interaction has been considered. Owing to the quasi-planar structure, first-principles calculations show that interlayer distances (the height difference between the downmost atoms in the upper layer and the topmost atoms in the lower layer) are 3.09 {\AA} and 3.37 {\AA} for the AA-8-$Pmnm$ and AB-16-$Amam$ borophenes, respectively. These distances are slightly shorter than the interlayer spacing in graphite, the corresponding calculated value of 3.45 {\AA} is in good agreement with the experimental value \cite{R38}. Hence, both the stacking sequence and the puckered structure of 8-$Pmmn$ borophene result in different interlayer distance. The AA-8-$Pmnm$ structure is 5 meV/atom lower in energy than the AB-16-$Amam$ structure. Note that the calculated energy difference between the bulk phases of $\alpha$-B$_{12}$ and $\beta$-B$_{106}$ is about 2 meV/atom \cite{R39}. Therefore, the stacking sequence play a decisive role in the relative stability of vdW-coupled stacked borophenes. Both the AA-8-$Pmnm$ and AB-16-$Amam$ borophene configurations are close in energy compared to other predicted three-dimensional (3D) boron phases (see Table \hyperref[t1]{I}), but are higher in energy relative to bulk $\alpha$-B$_{12}$, indicating these are metastable phases. While various structures of borophene have been reported, neither stacked nor fused borophenes have been explored extensively \cite{R40,R41}, even less for their high-pressure phases.
\label{f2}\label{f2}
\begin{figure}[t]
\begin{center}
\includegraphics[width=8.5cm]{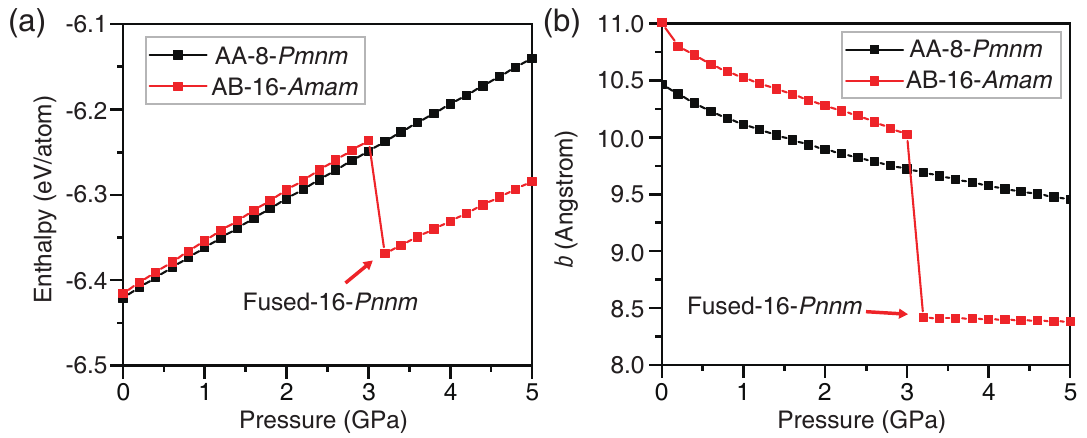}
\caption{%
(a) Enthalpy and (b) lattice parameter $b$ as a function of pressure.}
\end{center}
\end{figure}

\begin{table*}[t]\label{t1}
\caption{Structural parameters and the vdW-corrected total energies of the studied boron allotropes.}
\begin{tabular*}{14cm}{@{\extracolsep{\fill}}cccc}
\hline\hline
Structure                             & Lattice parameters                                          & Atomic position          & Total energy\\
(symmetry)                            &      ({\AA})                                                & (fractional coordinates) & (eV/atom)   \\
\hline
AA-8-$Pmnm$                           & $a$ = 3.25, $b$ = 5.27, $c$ = 4.51                          & B1(0.000,0.424,0.184)    & -6.420      \\
                                      & $\alpha$ = $\beta$ = $\gamma$ = 90$^{\circ}$                & B2(0.247,0.707,0.000)    &             \\
                                                                                                                                             \\
AB-16-$Amam$                          & $a$ = 3.25, $b$ = 11.10, $c$ = 4.51                         & B1(0.000, 0.152, 0.003)  & -6.415      \\
                                      & $\alpha$ = $\beta$ = $\gamma$ = 90$^{\circ}$                & B2(0.184, 0.286, 0.250)  &             \\
                                                                                                                                             \\
AB-16-$Pnnm$                          & $a$ = 3.20, $b$ = 8.48, $c$ = 4.50                          & B1(0.171, 0.074, 0.000)  & -6.519      \\
                                      & $\alpha$ = $\beta$ = $\gamma$ = 90$^{\circ}$                & B2(0.159, 0.352, 0.500)  &             \\
                                      &                                                             & B3(0.136, 0.690, 0.188)  &             \\
                                                                                                                                             \\
$\alpha$-B$_{12}$                     & $a$ = $b$ = $c$ = 5.03                                      & B1(0.010, 0.010, 0.654)  & -6.808      \\
($R$-3$m$)                            & $\alpha$ = $\beta$ = $\gamma$ = 58.0$^{\circ}$              & B2(0.222, 0.222, 0.630)  &             \\
                                                                                                                                             \\
8-$Pmmn$ \cite{R09}                    & $a$ = 3.25, $b$ = 4.52, $c$ = 13.00                         & B1(0.000, 0.316, 0.531)  & -6.395      \\
                                      & $\alpha$ = $\beta$ = $\gamma$ = 90$^{\circ}$                & B2(0.253, 0.000, 0.584)  &             \\
                                                                                                                                             \\
3D borophene \cite{R41}               & $a$ = 5.14, $b$ = 1.85, $c$ = 2.80                          & B1(0.167, 0.000, 0.000)  & -6.429      \\
($Immm$)                              & $\alpha$ = $\beta$ = $\gamma$ = 90$^{\circ}$                &                          &             \\
                                                                                                                                             \\
3D-$\alpha$$^\prime$ boron \cite{R50} & $a$ = 7.82, $b$ = 8.00, $c$ = 5.05                          & B1(0.000, 0.398, 0.568)  & -6.431      \\
($Cmcm$)                              & $\alpha$ = $\beta$ = $\gamma$ = 90$^{\circ}$                & B2(0.168, 0.289, 0.084)  &             \\
                                      &                                                             & B3(0.176, 0.304, 0.750)  &             \\
                                                                                                                                             \\
$\alpha$ sheet \cite{R51}             & $a$ = $b$ = 5.06, $c$ = 15.00                               & B1(0.000, 0.332, 0.500)  & -6.332      \\
($P$6/$mmm$)                          & $\alpha$ = $\beta$ = 90$^{\circ}$, $\gamma$ = 120$^{\circ}$ & B2(0.333, 0.667, 0.500)  &             \\
\hline\hline
\end{tabular*}
\end{table*}

Graphite can be converted into a novel form of diamond (cold compressed graphite) at high pressure or cubic diamond at high pressure and high temperature as a resulting of the $sp$$^{2}$-$sp$$^{3}$ transition \cite{R42,R43}. It is interesting to study the behavior of the stacked borophenes under pressure. The calculated pressure-enthalpy curves of the stacked borophene configurations are presented in Fig. \hyperref[f2]{2(a)}. One observes that the AA-8-$Pmnm$ phase is thermodynamically more stable than the AB-16-$Amam$ phase up to 2.75 GPa. Strikingly, the enthalpy of the AB-16-$Amam$ borophene decreases abruptly at 3 GPa, suggesting a structural phase transition. The variation of lattice parameter $b$ as a function of pressure is shown in Fig. \hyperref[f2]{2(b)}. This quantity also exhibits an abrupt decrease at the phase transition point for the AB-16-$Amam$ borophene. By analyzing the relaxed structure of the AB-16-$Amam$ phase at 3 GPa, we conclude that the phase transition takes place by overcoming the weak vdW interaction with bond breaking and bond reforming, leading to the connection of the adjacent layers. The relaxed structure is termed as the fused AB-16-$Pnnm$ borophene [Figs. \hyperref[f1]{1(c)} and \hyperref[f1]{1(f)}]. Compared to the AB-16-$Amam$ structure, the graphene-like sublattice in the fused AB-16-$Pnnm$ borophene is retained, while the boron chains in the adjacent layers are connected in a puckered arrangement. Although the interlayer distance in the AA-8-$Pmnm$ phase (3.09 {\AA}) is shorter than that in the AB-16-$Amam$ phase (3.37 {\AA}), as shown in Figs. \hyperref[f1]{1(a)} and \hyperref[f1]{1(b)}, the nearest B-B distance between the adjacent layers in the AA-8-$Pmnm$ phase (3.82 {\AA}) is longer than its interlayer distance (3.09 {\AA}), whereas it is the same for the AB-16-$Amam$ borophene (3.37 {\AA}). The longer B-B distance between adjacent layers in the AA-8-$Pmnm$ borophene seems to prevent the occurrence of direct phase transformation. Furthermore, the MD simulations showed the vdW-coupled borophenes were dynamically unstable and transformed to the fused phase under some conditions, whereas the fused AB-16-$Pnnm$ borophene was dynamically stable at least up to 1000 K. The dynamical stability of fused AB-16-$Pnnm$ borophene was also confirmed by its phonon dispersion curve \cite{R11}.

\begin{figure}[t]\label{f4}\label{f3}
\begin{center}
\includegraphics[width=8.0cm]{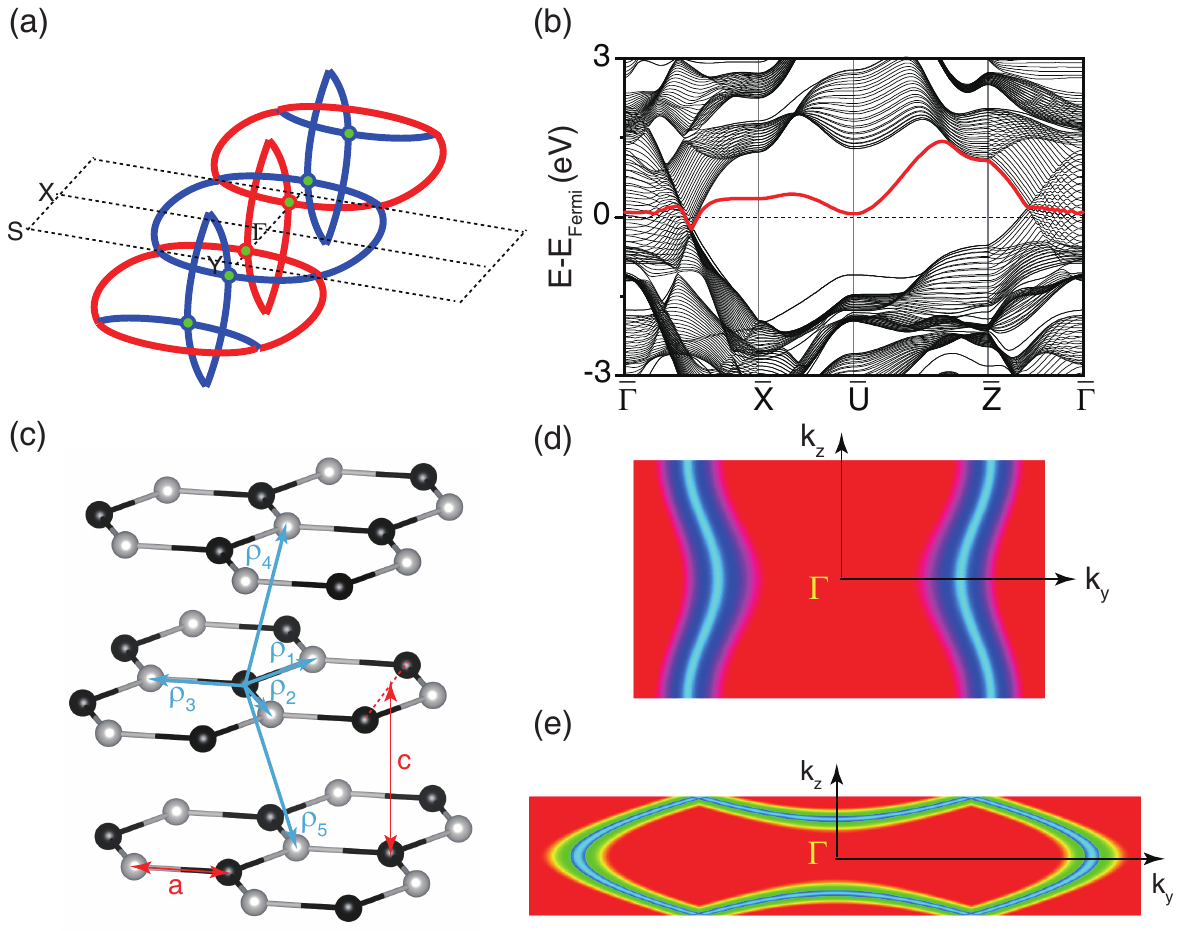}
\caption{%
(a) The distribution of the nodal lines of the fused AB-16-P$nnm$ borophene in the extended Brillouin zone. Two sets of nodal chains are colored in red and blue. The touching nodal points are colored in green. (b) The (010) surface states of the fused AB-16-$Pnnm$ borophene. (c) The graphene model adopted the same stacking sequence as AB-16-$Amam$ borophene, and two inequivalent sites are colored in black and gray, respectively. (d) and (e) The nodal line described by equation (1) when $J^{\prime}$ = -0.1$J$ and $J^{\prime}$ = -0.9$J$. The BZ plotted in (d) and (e) represent the different symmetries of AB-16-$Amam$ and fused AB-16-$Pnnm$ borophenes.}
\end{center}
\end{figure}

The 8-$Pmmn$ borophene hosts tilted and anisotropic massless Dirac fermion quasiparticles owing to its unique crystal structure. For the 2D 8-$Pmmn$ phase, as the interlayer distance is larger than 10 {\AA} (along $y$ direction), there is no interaction between adjacent layers [Figs. \hyperref[f3]{3(a)} and \hyperref[f3]{3(b)}]. As the interlayer distance is decreased to a certain value (e.g., 3.37 {\AA}), the AB-16-$Amam$ borophene is formed. The interlayer coupling transforms the Dirac points of 8-$Pmmn$ borophene into the nodal lines of AB-16-$Amam$ borophene [Figs. \hyperref[f3]{3(c)} and \hyperref[f3]{3(d)}]. As the interlayer distance is further decreased, AB-16-$Amam$ borophene may transform into the fused AB-16-$Pnnm$ phase under pressure (or under uniaxial compression). The mechanism underlying the band structure evolution is complex but intriguing. As shown in Fig. \hyperref[f1]{3(e)}, the orbital-resolved band structure shows that the fused AB-16-$Pnnm$ borophene is a semimetal with several crossing points at the vicinity of the Fermi level $E$$_{\rm F}$. Furthermore, the electrons at the crossing points 1 and 3 and the holes at the points 2 and 4 lead to nonzero density of states (DOS) at $E$$_{\rm F}$. The densities of the electrons and holes are evaluated by integrating their occupations of four bands near $E$$_{\rm F}$ \cite{R39}. The calculated electron and hole concentrations are equal to 1.76 $\times$ 10$^{20}$ cm$^{-3}$, which is less than the commonly accepted upper limit for semimetals (10$^{22}$ cm$^{-3}$). The coexistence of twofold and fourfold degenerate crossing points associated with the same concentration of electrons and holes suggests that the fused AB-16-$Pnnm$ borophene is a compensated topological semimetal (TSM). For the crossing point 1 at $\Gamma$-X segment, the Fermi velocities along the $k_{x}$ direction are 1.04 $\times$ 10$^{6}$ and 1.10 $\times$ 10$^{6}$ m/s, exceeding the calculated value of graphene (0.82 $\times$ 10$^{6}$ m/s) and being almost the same as the maximum value for 8-$Pmmn$ borophene (1.16 $\times$ 10$^{6}$ m/s) \cite{R09}. As a result, the AB-16-$Pnnm$ borophene partly inherits the Dirac band dispersion of the 8-$Pmmn$ structure in the $k_{x}$ direction, while the dispersion is different in the $k_{y}$-$k_{z}$ plane. Further band structure calculations show the nodal-line distribution of the fused AB-16-$Pnnm$ phase in the extended Brillouin zone (BZ) [Figs. \hyperref[f3]{3(f)} and \hyperref[f4]{4(a)}], consisting of overlapping nodal loops in the $k_{x}$-$k_{y}$ plane (by crossing the BZ boundary), while the nodal loops in the $k_{y}$-$k_{z}$ plane are separated. These two sets of nodal lines are orthogonal and intersect at the crossing point 3 [Fig. \hyperref[f3]{3(e)}]. According to the space group $Pnnm$, the point groups on the $\Gamma$XSY and $\Gamma$YTZ planes are both $C_{\rm s}$. The nodal loops in the $k_{x}$-$k_{y}$ plane are protected by the mirror operator $M_{z}$, whereas others in the $k_{y}$-$k_{z}$ plane are protected by the glide operator \{$M_{z}$$|$\textbf{\emph{a}}/2+\textbf{\emph{b}}/2+\textbf{\emph{c}}/2\} \cite{R11}. Overall, the fused AB-16-$Pnnm$ borophene can be classified as a nodal-chain semimetal. One of the most interesting features of the nodal-line materials is the presence of drumhead surface states, which may provide a route to higher-temperature superconductivity \cite{R44,R45}. The surface states of the fused AB-16-$Pnnm$ borophene are calculated using the fifteen-layer-thick (010) slab model. As shown in Fig. \hyperref[f4]{4(b)}, the drumhead-like surface state (colored in red) connects nodal points 1 and 2 in the $\overline{\rm \Gamma}$-$\overline{\rm X}$-$\overline{\rm U}$-$\overline{\rm Z}$-$\overline{\rm \Gamma}$ path. The dispersion is flat around the high-symmetry point $\overline{\rm X}$, which is expected to facilitate the detection in the future experiment. We would like to point out that spin-orbit coupling (SOC) can break the symmetry and open a band gap at the crossing point. However, due to the very weak SOC in the studied light-element systems, only a tiny gap opens at the nodal points, e.g. $\sim$1.35 meV at the fourfold degenerate point, which is too weak to affect the semimetal properties.

\begin{figure}[t]
\begin{center}
\includegraphics[width=8.0cm]{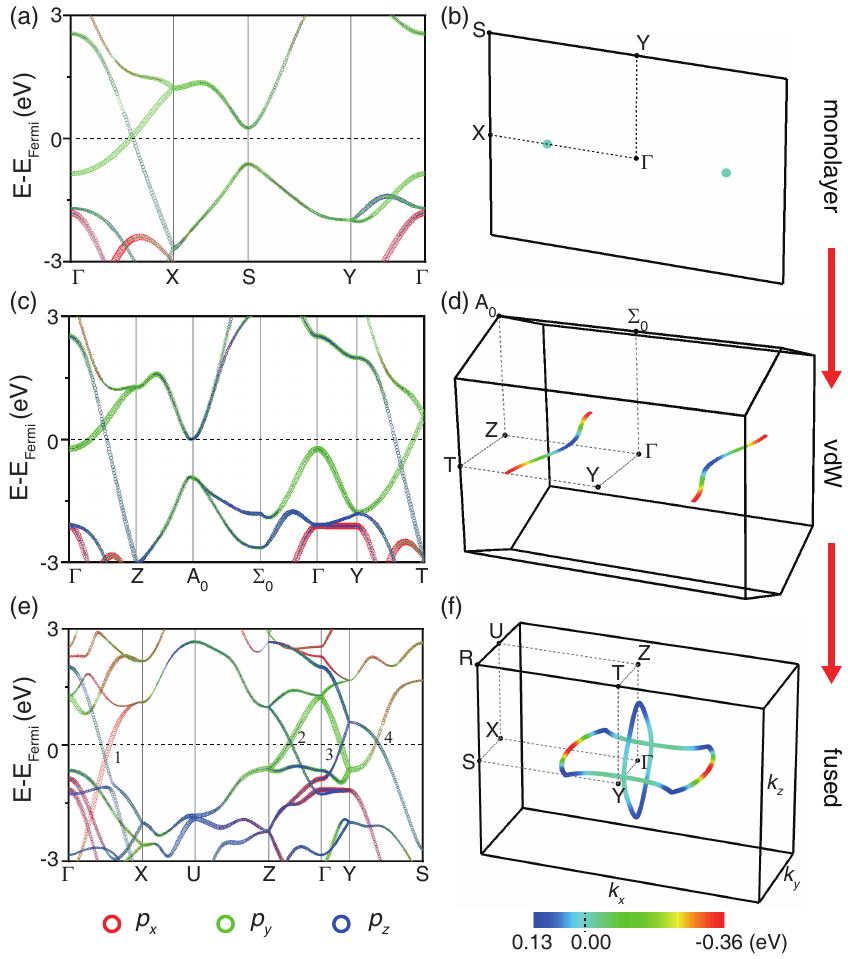}
\caption{%
(a, c, e) Band structures of the 8-$Pmmn$, AB-16-$Amam$, and the fused AB-16-$Pnnm$ phases, respectively. Red, green, and blue colors indicate the weight of $p_{x}$, $p_{y}$, $p_{z}$ orbitals, respectively. (b, d, f) The corresponding distribution of the crossing points for (a), (c), (e).}
\end{center}
\end{figure}

Since the graphene-like substructure of 8-$Pmmn$ borophene is the key factor for the emergence of the Dirac cone, a simple tight-binding (TB) model derived from graphene is constructed to qualitatively analyze the electronic evolution from Dirac semimetal to nodal-line semimetal  after stacking 8-$Pmmn$ borophene to form 3D structures. As shown in Fig. \hyperref[f4]{4(c)}, the nearest intralayer and interlayer hopping vectors from two inequivalent sites were considered as:$$ \boldsymbol{\rho_{1}} = \frac{a}{2}\hat{x} + \frac{\sqrt{3}a}{2}\hat{y} \quad \boldsymbol{\rho_{2}} = \frac{a}{2}\hat{x} - \frac{\sqrt{3}a}{2}\hat{y} \quad \boldsymbol{\rho_{3}} = -a\hat{x} $$ $$
\boldsymbol{\rho_{4}} = \frac{a}{2}\hat{x} + c\hat{z} \quad \boldsymbol{\rho_{5}} = \frac{a}{2}\hat{x} - c\hat{z} $$ The TB Hamiltonian based on the $p_{z}$ orbital was constructed to reveal the band structure in proximity of the $E$$_{\rm F}$ as:
\begin{equation}
\begin{aligned}
 &E(k) = \pm [2J^{\prime2}(1+\cos(2ck_{z}))
  \\ &+J^{2}(3+2\cos(\sqrt{3}k_{y}a)+ 4\cos(\frac{3}{2}a(k_{x}-\frac{2\pi}{3a}))\cos(\frac{3}{2}k_{y}a))
  \\&+4JJ^{\prime}\cos(ck_{z})(2\cos(\frac{\sqrt{3}}{2}ak_{y})+\cos(\frac{3}{2}a(k_{x}-\frac{2\pi}{3a})))]^{0.5} \
\end{aligned}
\end{equation}
Where the parameters of $J$, $J^{\prime}$, $a$, and $c$ represent the intralayer hopping strength, interlayer hopping strength,  C-C bond length and interlayer distance, respectively. Assuming $J$\textgreater0 and plotting the band structure at plane of $k_{x}$=0, the electronic transition arised from Dirac semimetal to nodal-line semimetal when -0.5$J$\textless$J^{\prime}$\textless0, which is correspond to stack the 8-$Pmmm$ borophene into the AB-16-$Amam$ borophene [Figs. \hyperref[f3]{3(d)} and \hyperref[f4]{4(d)}]. Similarly, the electronic transition is transformed from nodal-line semimetal to nodal-loop or nodal-chain semimetal when $J^{\prime}$\textless-0.5$J$. For instance, the phase transition from the AB-16-$Amam$-borophene to the fused AB-16-$Pnnm$ borophene [Figs. \hyperref[f3]{3(f)} and \hyperref[f4]{4(e)}]. Reducing the value of $J^{\prime}$ indicates the enhancement of interlayer interaction. Note that the interlayer interaction of $J^{\prime}$\textless-0.5$J$ is prohibited in the stacked graphene system. The interlayer interaction is too strong to break the graphene sheets, leading to the phase transition from graphite to diamond, while it can be realized in the stacked 8-$Pmmm$ borophene system. As long as the graphene-like substructure is preserved, the essence of topological semimetal is retained even there is a partially bond breaking and reforming during phase transition.

In conclusion, we investigated the stacked 8-$Pmmn$ borophene in different forms. Under increasing pressure, the interatomic distances typically decreased. The valence and conduction bands are thus expected to broaden, leading to the pressure-induced metallization \cite{R46}. The AB-stacked borophene is transformed into the fused borophene at $\sim$3 GPa associated with the bond breaking and reforming between the adjacent boron chains. However, due to the preserved graphene-like substructure, the pressure-induced semimetal-semimetal transition takes place in the stacked 8-$Pmmn$ borophene---that is, a nodal-line semimetal (AB-16-$Amam$ phase) transforms to a nodal-chain semimetal (fused AB-16-$Pnnm$ phase), which is different from the common semimetal-metal (semimetal-semiconductor) transition. Recently, bilayer borophenes were successfully synthesized on the Ag(111) and Cu(111) substrates \cite{R47,R48}. Meanwhile, the 8-$Pmmn$ borophene was predicted to be grown on the metal substrates because its lattice constants match with the (110) surface of several metals and metal oxides \cite{R09,R49}. It is therefore anticipated that layer-stacked borophenes might be synthesized in the near future. If the synthesis succeeds, new bulk allotropes of boron could be formed under pressure and may be quenchable to ambient condition. These would extremely expand the phase diagram of elemental boron.

This work was supported by the National Natural Science Foundation of China (Grants No. 52025026, 11874224, 52090020). Q.W. and O.V.Y. acknowledge support by the NCCR Marvel. The calculation was performed on the TianheII supercomputer at Chinese National Supercomputer Center in Guangzhou and at the Swiss National Supercomputing Centre (CSCS) under Project No. s1008.



\end{document}